# Observation of enhanced exchange bias behavior in NiCoMnSb Heusler alloys

**Ajaya K. Nayak[1], K. G. Suresh[1] and A. K. Nigam[2]**

[1]Magnetic Materials Laboratory, Department of Physics, Indian Institute of Technology Bombay, Mumbai-400076, India.

[2]Tata Institute of Fundamental Research, Homi Bhabha Road, Mumbai-400005, India

E-mail: suresh@phy.iitb.ac.in

*Abstract*

*We report the observation of large exchange bias in $Ni_{50-x}Co_xMn_{38}Sb_{12}$ Heusler alloys with x=0, 2, 3, 4, 5, which is attributed to the coexistence of ferromagnetic and antiferromagnetic phases in the martensitic phase. The phase coexistence is possibly due to the supercooling of the high temperature ferromagnetic phase and the predominant antiferromagnetic component in the martensitic phase. The presence of exchange bias is well supported by the observation of training effect. The exchange bias field increases with Co concentration. The maximum value of 480 Oe at T=3K is observed in x=5 after field cooling in 50 kOe, which is almost double the highest value reported so far in any Heusler alloy system. Increase in the antiferromagnetic coupling after Co substitution is found to be responsible for the increase in the exchange bias.*

## 1. Introduction

After the discovery of exchange bias (EB) behavior in Co particle embedded with its oxide by Meiklejohn and Bean [1], many systems have been studied because of their potential applications in many devices such as sensors, magnetic recording media, read heads etc [2]. This extensive research has resulted in the discovery of many EB materials, including thin films and bulk inhomogeneous materials [2-8]. The EB behavior has been observed in materials having ferromagnetic (FM)- antiferromagnetic (AFM) interface. When these materials are cooled below the Neel temperature ($T_N$) of the AFM phase, unidirectional anisotropy arises, resulting in exchange bias.

Ni-Mn-X (X=Ga, Sn, In, Sb) ferromagnetic Heusler alloys have attracted great research interest owing to their multifunctional properties such as shape memory effect, magnetocaloric effect and magnetoresistance which are associated with the first order austenite to martensitic transition [9-13]. Recently exchange bias behavior has been reported for some of these alloys [14-16]. In present work we have studied the exchange bias behavior on partial substitution of Co for Ni in $Ni_{50}Mn_{38}Sb_{12}$. We find that much larger EB can be achieved with Co substitution in the off-stoichiometric NiMnSb alloys.

## 2. Experimental Details

The method of preparation of the alloys is reported elsewhere [13]. The structural characterization was carried out by powder x-ray diffractograms (XRD) using Cu-Kα radiation. The magnetization measurements were carried out using a vibrating sample magnetometer attached to a Physical Property Measurement System (Quantum Design, PPMS-6500)/ SQUID magnetometer (Quantum Design). The temperature dependence of magnetization ($M$) measurements has been performed in three different modes. In the zero field cooled (ZFC) mode, the sample was initially cooled to 5 K without applying any magnetic field and then data was taken as the temperature was increased from 5 K by applying field of 1 kOe . In the field cooled cooling (FCC) mode, the data was collected during the cooling with field, while in the filed cooled warming (FCW) mode, the data was collected while heating, after field cooling.

## 3. Results and discussion

The XRD patterns for $Ni_{50-x}Co_xMn_{38}Sb_{12}$ (figure 1) taken at room temperature shows the compounds with x≤3 are martensitic (orthorhombic), whereas the ones with x=4 and 5 possess austenite (cubic) phase at room temperature. But in the case of x=4 the existence of small peak near the (220) austenite peak indicates the presence of very small amount of martensitic phase, as this phase starts just below the room temperature. However, the Rietveld refinement for x=5 shows that it is completely in the austenite phase at room temperature. It has also been found that the variation of the lattice parameters with Co concentration is found to be quite negligible.

Figure 2 shows the temperature dependence of magnetization in $Ni_{50-x}Co_xMn_{38}Sb_{12}$ alloys with x=2, 3, 4 and 5. For all the four compounds ferromagnetic-paramagnetic transition of the austenite phase ($T_C^A$) occurs nearly around 330 K. Below $T_C^A$ the magnetization increases to a maximum value at the martensitic start temperature ($M_S$) and then attains the minimum at the martensitic finish temperature ($M_F$). Below, $M_F$, the magnetization shows an increase, corresponding to the magnetic ordering temperature of the martensitic phase ($T_C^M$). It is observed that the $T_C^M$ of the parent compound (x=0) is in good agreement with the result reported previously [17]. It can be seen that the martensitic transition temperatures ($M_S$ or $M_F$) decreases monotonically with Co concentration. The $T_C^M$ also decreases monotonically with Co, as is evident from the figure 2. For example $T_C^M$ for x=0 is 280K and becomes 235K for x=5. The magnetic and martensitic transition temperatures are indicated only in the figure for x=4 and 5, for better clarity. In all the compounds, the martensitic transition is of first order in nature, as revealed by the thermal hysteresis between the FCC and FCW data.

The sharp decrease in magnetization associated with the first order austenite-martensite transition (over $M_S$-$M_F$ regime) indicates the presence of some antiferromagnetic component in the martensitic phase. Using extended x-ray absorption fine-structure measurements in NiMnSn, it has been shown that the Mn-Mn distance between the Mn atoms occupying the Mn and Sn sublattices decreases under transformation from austenite phase to martensitic phase, thereby introducing an antiferromagnetic component [18]. It has been suggested by these authors that while the coupling within between the regular Mn sites is ferromagnetic, the one between Mn and Sb sites (containing Mn) is antiferromagnetic. It is quite likely that such a scenario occurs in the present in NiMnSb as well. At low temperatures (below $T_b$, shown in Figure 2) the drop in the ZFC magnetization clearly indicates the presence of AFM in the martensitic phase. However, the FCW magnetization resembles that of a ferromagnet. There is no considerable change in the overall features of the M-T plot as a function of Co concentration, except that the transition temperatures are shifted.

Figure 3 shows the isothermal magnetization curves of $Ni_{45}Co_5Mn_{38}Sb_{12}$ at different temperatures, after cooling in a field of 50 kOe from 350 K. After reaching the desired temperature, isothermal magnetization data has been taken in the range of ± 20 kOe. For better clarity, the hysteresis loops are shown only in the range of ± 5 kOe in the figure. As is evident from figure 3, the hysteresis loop at 3 K has shifted completely to the negative field axis. The amount of the shifting of the hysteresis loop decreases with temperature. The hysteresis loop is found to be symmetric at temperatures close to 80 K, which nearly coincides with the blocking temperature ($T_b$) observed from the M-T data shown in figure 2. The presence of shifted loops points to the fact that there is exchange bias present in the system due to the existence of FM/AFM mixed phase at low temperatures. As the M-T data has clearly shown, there is AFM phase at low temperatures. Due to the complex magnetic structure associated with the martensitic phase, there is an FM component as well in the martensitic region. In addition, as reported in martensitic systems such as doped $CeFe_2$ [19], the first order transition causes the supercooling of the high temperature FM phase (austenite), which coexists with the predominant AFM phase in the martensitic regime. These factors must be responsible for the creation of the FM-AFM exchange coupling.

Fig. 4 shows variation of EB field ($H_{EB}$) and coercivity ($H_C$) as a function of temperature for $Ni_{45}Co_5Mn_{38}Sb_{12}$. $H_{EB}$ and $H_C$ are calculated using $H_{EB} = -(H_1 + H_2)/2$ and $H_C = |H_1 - H_2|/2$, where $H_1$ and $H_2$ are the lower and upper cut-off fields at which the magnetization becomes zero. The $H_{EB}$ shows an approximately linear decrease with temperature and becomes zero above 80 K. On the other hand, $H_C$ initially increases with temperature and then decreases after reaching a maximum value. Inset of Figure 3 shows the variation of exchange bias field with Co concentration at 3 K, which shows that the $H_{EB}$ increases from 245 Oe to 480 Oe, as $x$ is increased from 0 to 5. Therefore, it is quite evident that the FM/AFM coupling gets reinforced with Co. It is of importance to note that the highest $H_{EB}$ value ever reported under identical experimental conditions for Heusler alloys is 248 Oe, i.e., in the case of $Ni_{50}Mn_{38.5}Sb_{11.5}$ [15].

Figure 5 shows the cooling field dependence of $H_{EB}$ and $H_C$ in $Ni_{45}Co_5Mn_{38}Sb_{12}$ at 5 K. The sample was cooled from 350 K to 5 K in different fields. The magnetic isotherms were measured in ±20 kOe range. It shows that $H_{EB}$ decreases approximately linearly with cooling field and saturates in higher cooling fields whereas $H_C$ increases with cooling field. Since in low cooling field all the FM regions are not fully saturated, the alignment of the FM region increases along a certain direction and hence the interaction at the interface increases due to the exchange coupling, giving rise to large $H_{EB}$. The decrease in $H_{EB}$ is due to growth of FM clusters in high cooling field, as at high cooling field the alignment of the moments of FM clusters as well as the size of FM clusters increases. As the FM clusters grow up, exchange bias is reduced, which is seen in FM/AFM thin films where exchange bias is inversely proportional to the thickness of the FM layer [2].

An evidence for the presence of AFM-FM coexistence can be seen in Figure 6 which shows the ZFC magnetization isotherm, showing the double shifted loop, similar to the one seen in EB systems [15]. This is due to the division of AFM region into two opposing alignments with respect to the local FM component and during the magnetization measurement each of these regions couple oppositely with the adjacent FM region, giving rise to the double shifted loop.

The most striking property shown by most of the EB materials is the training effect, which reveals the decrease of $H_{EB}$ by cycling the system through several hysteresis loops [2]. To verify whether such an effect is present in this case, we have measured several hysteresis loops at 5 K after field cooling in 40 kOe. The first and fourth loops are shown for $Ni_{45}Co_5Mn_{38}Sb_{12}$ in Figure 7. Only the middle region of the curve is shown for clear visibility. It confirms that training effect is present in this system. The inset of Figure 6 shows the dependence of $H_{EB}$ on the number of field cycles ($n$), which shows that $H_{EB}$ decreases with $n$. Generally, the variation of $H_{EB}$ with $n$ is given by the relation

$$H_{EB} - H_{EB\infty} \propto \sqrt{n} \qquad (1)$$

where $H_{EB\infty}$ is the EB field in the infinite $n$ limit [7]. The fit of the equation 1, shown in the inset of Figure 7, shows that this variation is indeed followed very well by the present system. The value of $H_{EB\infty}$ is found to be 412 Oe. Training effect arises mainly because the ferromagnetic layer does not reverse homogenously during the magnetization reversal, possibly due to the rearrangement of AFM domain structure in the mixed AFM/FM phase [20]. This results in a partial loss of the net magnetization of antiferromagnetic layer during consecutive field cycles, leading to the reduction of EB. It is to be mentioned here that the training effect seen in $Ni_{45}Co_5Mn_{38}Sb_{12}$ is considerably larger than that observed in NiMnIn [16]. The anisotropy associated with Co may be responsible for this difference.

The main features seen from the present study are (i) the decrease in $H_{EB}$ with temperature and cooling field and (ii) the increase in $H_{EB}$ with Co concentration (shown in the inset of figure 4). The suppression of the AFM phase due to the increase in the temperature and the cooling field, which weakens the coupling at the FM-AFM interface explains the first observation. The most important feature seen from the present study is the increase in $H_{EB}$ with Co concentration, which suggests that there is an enhancement of the FM-AFM coupling with Co. One possibility is that the Co ions, which occupy the Ni site in austenite phase, may partially occupy the Sb site when the material undergoes the martensitic transition. Due to the ionic size difference between Sb and Co, this may effectively enhance the AFM between the Mn ions occupying the Mn and Sb sites. The fact that the $T_C^M$ decreases monotonically with Co concentration also confirms the increase in the AFM strength. Another reflection of this increase is seen in the low temperature magnetization taken after field cooling, which shows a decreasing trend, as shown in Figure 8 (a). It can be clearly seen that there is a significant reduction in the saturation magnetization after Co substitution, though a systematic trend with Co concentration is absent. The fact that the Co magnetic moment is larger than the Ni moment also plays a role in determining the net magnetization of the alloy. In the austenite phase, on the other hand, there is a monotonic increase in the magnetization with Co, as shown in figure 8 (b). The difference in these two trends in the austenite and martensitic phases lends some credence to our suggestion that the AFM strength increases with Co in the martensitic phase. At this point, it should also be kept in mind that there will be local anisotropy variations due to the small amount of Co substitution, which will influence the magnetization behavior associated with the FM component as well as the FM/AFM interface coupling. A detailed neutron diffraction study is essential to establish some of the above propositions regarding the magnetic state of these multifunctional materials.

## 4. Conclusions

In conclusion, we have observed large exchange bias in Co substituted Ni-Mn-Sb Heusler alloys below their martensitic transition temperature. EB arises due to the unidirectional anisotropy caused by the coupling between the FM and AFM at their interface. $H_{EB}$ is found to increase with Co concentration and a large value of 480 Oe is achieved in the sample with the highest Co concentration (x=5). The enhancement of the exchange bias with Co is attributed to the increase in the AFM coupling, which also manifests in the decrease in the low temperature magnetization and $T_C^M$. The observations of training effect and double shifted hysteresis loops strongly support the EB phenomenon in these materials. Therefore, the present results seem to be very promising in the search for multifunctional materials from the Heusler family.

Acknowledgement

The authors (KGS and AKN) thank BRNS, Govt. of India for funding this work.

**Figure captions**

Figure 1. Temperature variation of x-ray diffractograms of $Ni_{50-x}Co_xMn_{38}Sb_{12}$ compounds with x=0, 3, 4, 5. Rietveld refinement for x=5 showing single phase.

Figure 2. ZFC, FCC and FCW magnetization data as a function of temperature for $Ni_{50-x}Co_xMn_{38}Sb_{12}$ alloys with x=2, 3, 4, 5 in a field of 1 kOe.

Figure 3. M vs. H loops measured at various temperatures for $Ni_{45}Co_5Mn_{38}Sb_{12}$

Figure 4. Variation of exchange bias field ($H_{EB}$) and coercivity ($H_C$) with temperature for $Ni_{45}Co_5Mn_{38}Sb_{12}$. The inset shows the variation of $H_{EB}$ with Co concentration at 3K.

Figure 5. Variations of exchange bias field and coercivity with the cooling field for $Ni_{45}Co_5Mn_{38}Sb_{12}$ at 5 K.

Figure 6. Magnetic hysteresis loop obtained in the ZFC mode for $Ni_{45}Co_5Mn_{38}Sb_{12}$ at 3K.

Figure 7. Training effect of EB in $Ni_{45}Co_5Mn_{38}Sb_{12}$. The first and fourth loops are shown at 5 K after field cooling in 40 kOe. The inset shows the dependence of $H_{EB}$ on the number of field cycles (*n*). The solid line in the inset shows the best fit to equation 1.

Figure 8. *M-H* isotherms of $Ni_{50-x}Co_xMn_{38}Sb_{12}$ for x=0, 2, 3, 4 and 5 (a ) at 3K and (b) just above the martensitic start temperature

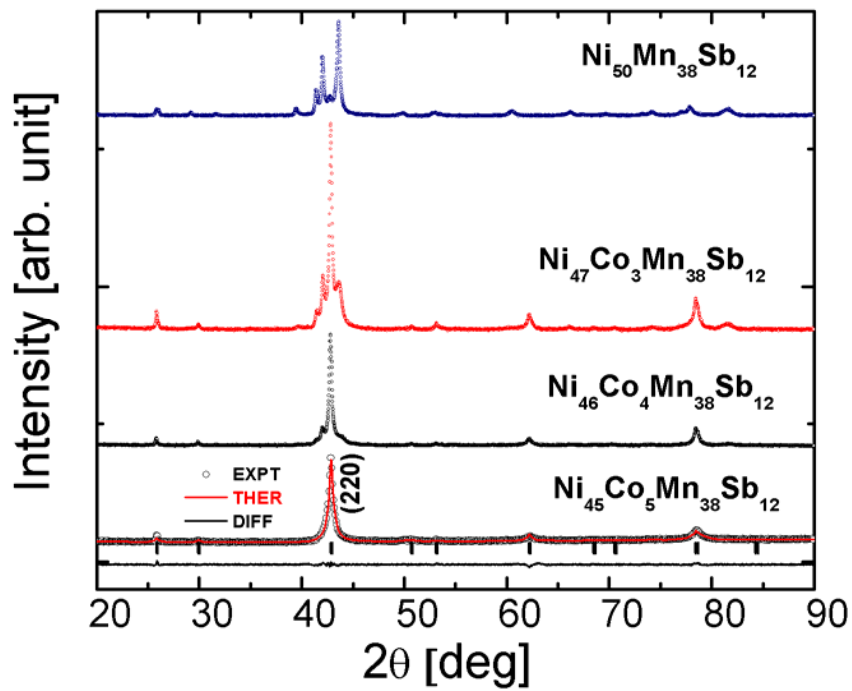
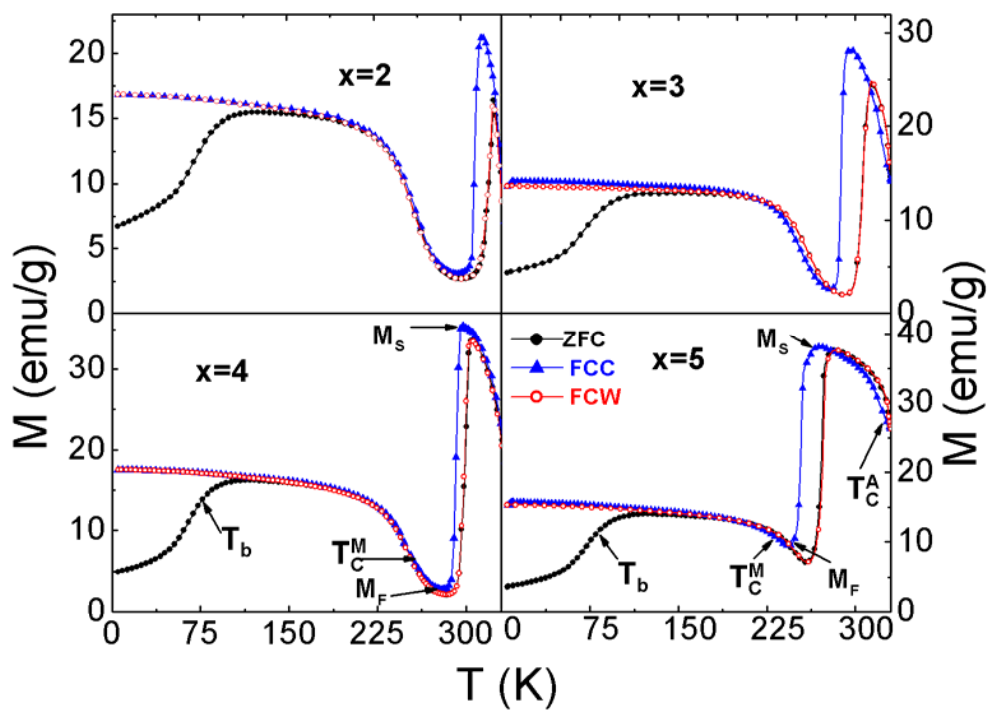

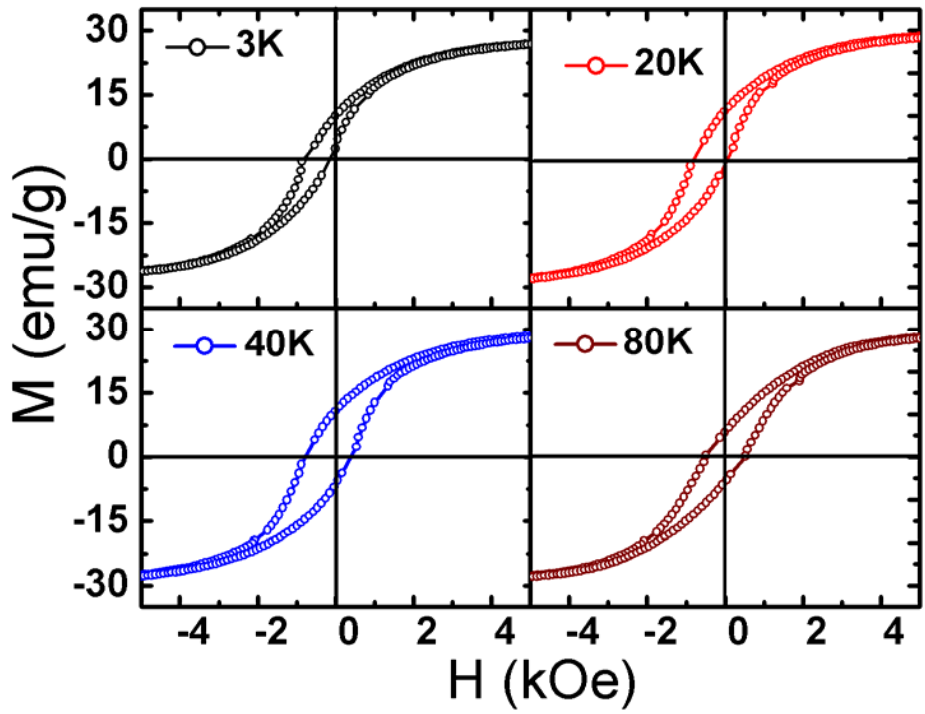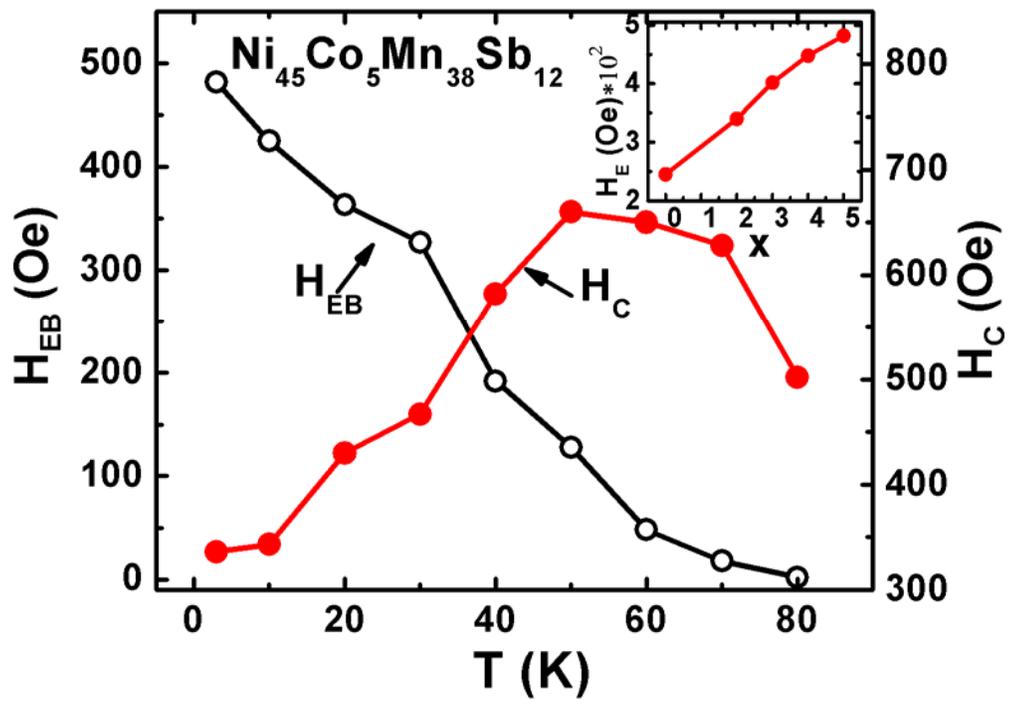

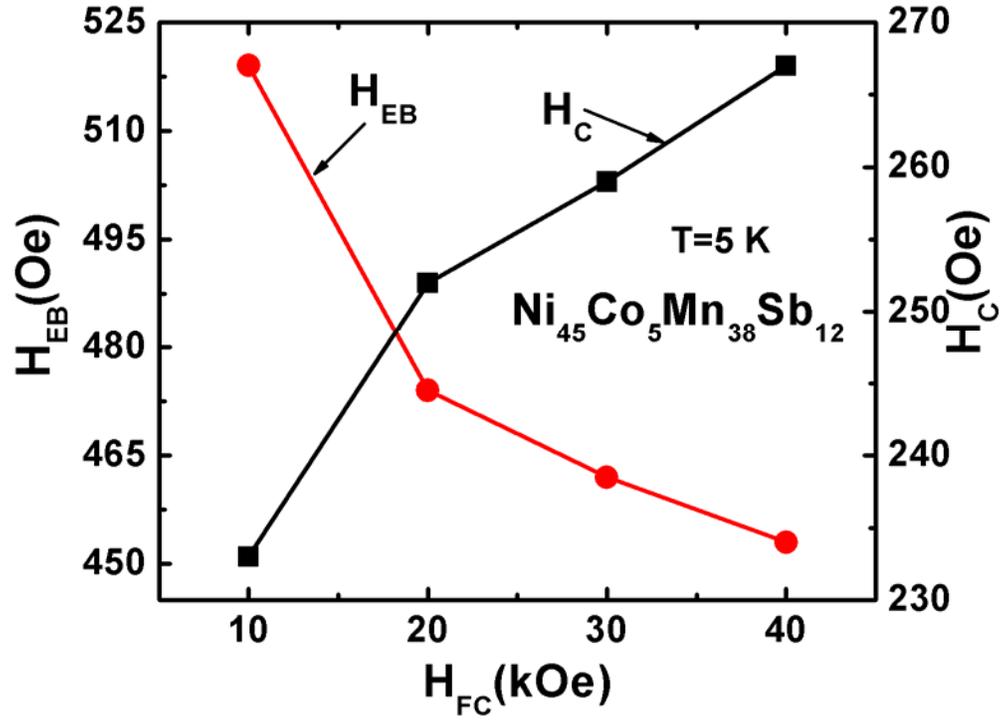

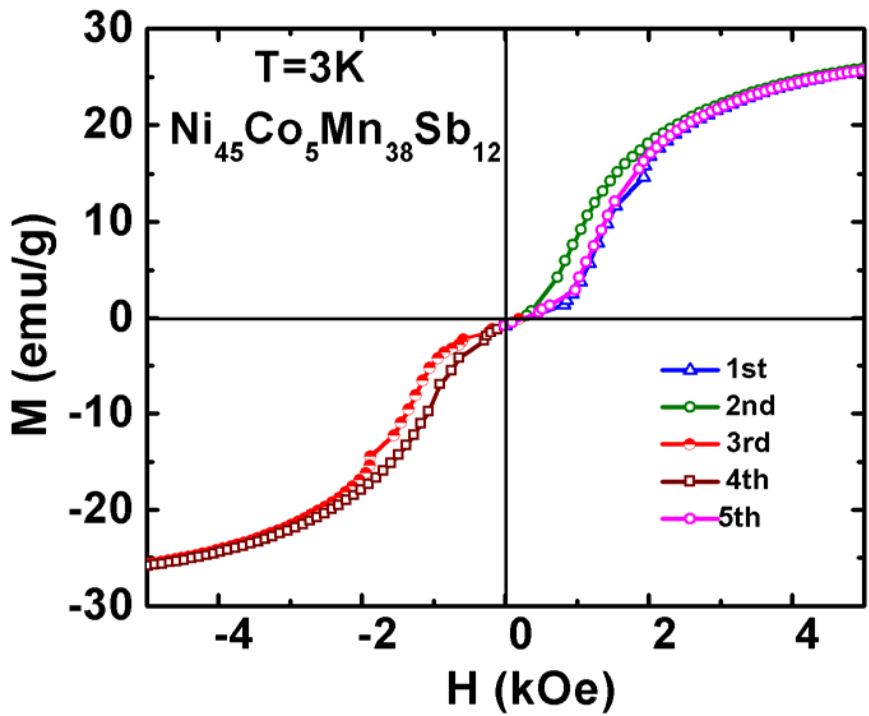

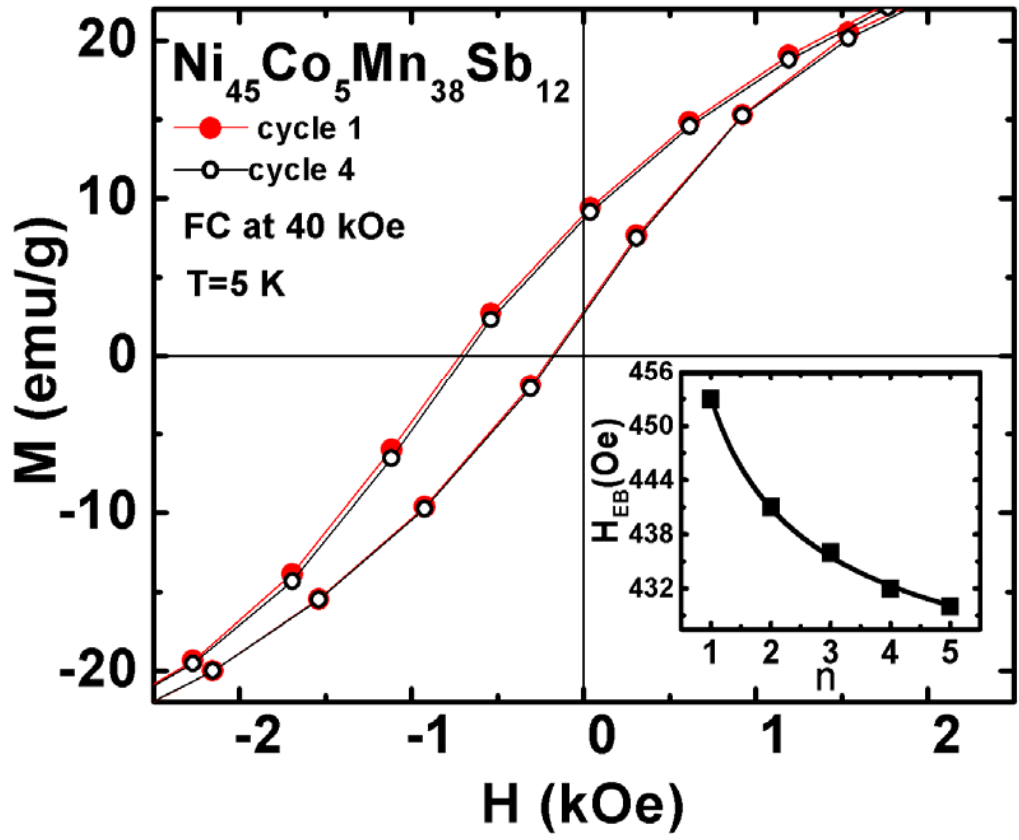

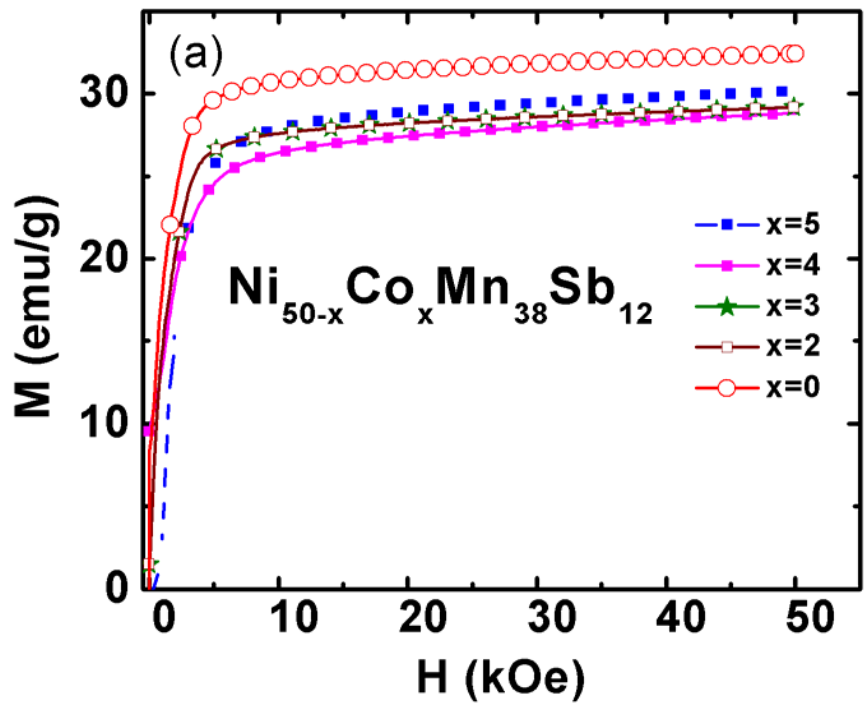

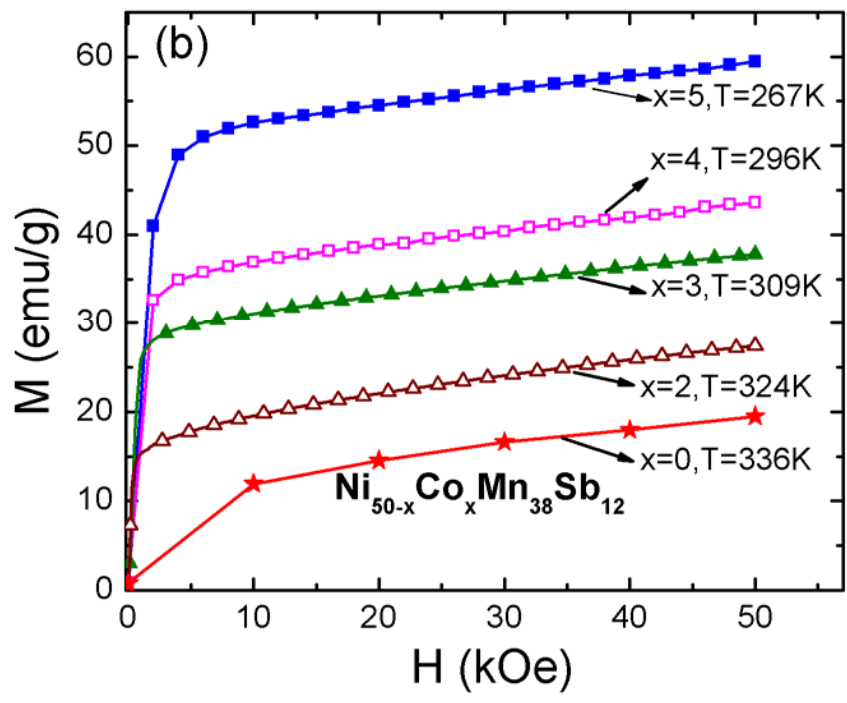